\documentclass[11pt]{article} \usepackage{hyperref} \pdfoutput=1

\begin{document} 
\title{The Life of a Vortex Knot}

\author{Dustin Kleckner\\ Martin Scheeler\\ William T. M. Irvine \\ \\\vspace{6pt} James Franck Institute and Department of Physics,\\ The University of Chicago, Chicago, Illinois 60637, USA}
\maketitle

\begin{abstract}
The idea that the knottedness (hydrodynamic Helicity) of a fluid flow is conserved has a long history in fluid mechanics.
The quintessential example of a knotted flow is a knotted vortex filament, however, owing to experimental difficulties, it has not been possible until recently to directly generate knotted vortices in real fluids.
Using 3D printed hydrofoils and high-speed laser scanning tomography, we generate vortex knots and links and measure their subsequent evolution.
In both cases, we find that the vortices deform and stretch  until a series of vortex reconnections occurs, eventually resulting several disjoint vortex rings.

This article accompanies a fluid dynamics video entered into the Gallery of Fluid Motion at the 66th Annual Meeting of the APS Division of Fluid Dynamics. 
\end{abstract}

Smoke rings are an everyday example of filamental vortex loops that are found in many complex flows.
Vortex loops with non-trivial topology -- linked or knotted together -- should be conserved in ideal fluids, as first noted by Lord Kelvin\cite{Kelvin1867}.
In real fluids, even superfluids, the situation is more complicated, and linking can change through reconnection events, whose detailed dynamics are challenging to resolve in both theory and experiment.
As a result, the dynamics of knots in non-ideal fluids is largely unresolved.

Previous methods for generating vortex loops are difficult or impossible to generalize to creating vortex links and knots, so we developed a new technique based on accelerating specially shaped hydrofoils\cite{Kleckner2013}.
Using this method, we were able to generate linked and knotted vortices for the first time, and visual their cores using bouyant micro-bubbles which are trapped in the regions of high vorticity.
Typically we generate vortices with an overall scale of $\sim$10 cm, corresponding to a Reynolds number of $Re \approx 10^4 - 10^5$.

To measure their three-dimensional evolution, we  developed an ultra-high speed laser scanning tomography apparatus.
By sweeping a laser sheet rapidly over the sample and imaging the scattered light with a high speed camera, we are able to reconstruct the bubble density (and vortex path) at 180 volumes per second.
We observe that topologically non-trivial vortices stretch and deform until vortex reconnections occur, ultimately resulting in unlinked rings after $\sim 0.5$ s.
We do not observe overall stretching or vortex reconnections in unlinked rings generated using the same method, even when significantly distorted, indicating that the topology is driving the reconnection dynamics.

These observations of linked and knotted vortices provide a first glimpse at their dynamics in experiment, raising many questions:
Are there all linked or knotted vortex configuration intrinsically unstable?
Is Helicity conserved through a reconnection event by changing bulk knotting to core twist?
Reconnections are also seen in superfluid vortices and magnetic field lines in plasmas; is there a universal topological mechanism that governs all of these events?


 \end{document}